\documentclass[fleqn,usenatbib]{mnras}

\usepackage{newtxtext,newtxmath}

\usepackage[T1]{fontenc}

\DeclareRobustCommand{\VAN}[3]{#2}
\let\VANthebibliography\thebibliography
\def\thebibliography{\DeclareRobustCommand{\VAN}[3]{##3}\VANthebibliography}

\usepackage{graphicx}	
\usepackage{amsmath}	
\usepackage{wasysym}    
\usepackage{xspace}
\makeatletter 
  \patchcmd{\NAT@citex}
    {\@citea\NAT@hyper@{%
      \NAT@nmfmt{\NAT@nm}%
      \hyper@natlinkbreak{\NAT@aysep\NAT@spacechar}{\@citeb\@extra@b@citeb}%
      \NAT@date}}
    {\@citea\NAT@nmfmt{\NAT@nm}%
    \NAT@aysep\NAT@spacechar\NAT@hyper@{\NAT@date}}{}{}

  \patchcmd{\NAT@citex}
    {\@citea\NAT@hyper@{%
      \NAT@nmfmt{\NAT@nm}%
      \hyper@natlinkbreak{\NAT@spacechar\NAT@@open\if*#1*\else#1\NAT@spacechar\fi}%
        {\@citeb\@extra@b@citeb}%
      \NAT@date}}
    {\@citea\NAT@nmfmt{\NAT@nm}%
    \NAT@spacechar\NAT@@open\if*#1*\else#1\NAT@spacechar\fi\NAT@hyper@{\NAT@date}}
    {}{}
\makeatother

\newcommand\Msun{\text{M}_{\astrosun}} 
\newcommand\Zsun{\text{Z}_{\astrosun}} 
\newcommand{\thesan}{\mbox{\textsc{THESAN}}\xspace}
\newcommand{\tngs}{\mbox{\textsc{TNG50}}\xspace}
\newcommand{\tngl}{\mbox{\textsc{TNG300}}\xspace}
\newcommand{\mtng}{\mbox{\textsc{MTNG}}\xspace}
\newcommand{\jwst}{\textit{JWST}\xspace}
\newcommand{\hst}{\textit{HST}\xspace}
\newcommand{\spitzer}{\textit{Spitzer}\xspace}

\title[MillenniumTNG -- Galaxy population at $z\geq8$]{The MillenniumTNG Project: The galaxy population at $z\geq8$}

\author[R. Kannan et al.]{%
Rahul Kannan,$^{1,2}$\thanks{E-mail: kannanr@yorku.ca}
Volker Springel,$^{3}$ Lars Hernquist,$^{1}$ R\"udiger Pakmor,$^{3}$ Ana Maria Delgado,$^{1}$%
\newauthor%
Boryana Hadzhiyska,$^{1}$ C\'esar Hern\'andez-Aguayo,$^{3}$ Monica Barrera,$^{3}$ Fulvio Ferlito,$^{3}$ Sownak Bose,$^{4}$ 
\newauthor%
Simon D. M. White,$^{3}$ Carlos Frenk,$^{4}$ Aaron Smith$^{1}$ and Enrico Garaldi$^{3}$
\\%
\\%
$^{1}$Center for Astrophysics | Harvard $\&$ Smithsonian, 60 Garden Street, Cambridge, MA 02138, USA\\%
$^{2}$Department of Physics and Astronomy, York University, 4700 Keele Street, Toronto, Ontario MJ3 1P3, Canada\\%
$^{2}$Max-Planck Institute for Astrophysics, Karl-Schwarzschild-Str.~1, D-85741 Garching, Germany\\%
$^{3}$Institute for Computational Cosmology, Department of Physics, Durham University, South Road, Durham, DH1 3LE, UK
}

\date{Accepted XXX. Received YYY; in original form ZZZ}

\pubyear{2022}

\begin{document}
\label{firstpage}
\pagerange{\pageref{firstpage}--\pageref{lastpage}}
\maketitle

\begin{abstract}
The early release science results from \jwst have yielded an unexpected abundance of high-redshift luminous galaxies that seems to be in tension with current theories of galaxy formation. However, it is currently difficult to draw definitive conclusions form these results as the sources have not yet been spectroscopically confirmed. It is in any case important to establish baseline predictions from current state-of-the-art galaxy formation models that can be compared and contrasted with these new measurements. In this work, we use the new large-volume \mbox{($L_\mathrm{box}\sim 740 \, \mathrm{cMpc}$)} hydrodynamic simulation of the MillenniumTNG project, suitably scaled to match results from higher resolution - smaller volume simulations, to make predictions for the high-redshift ($z\gtrsim8$) galaxy population and compare them to recent \textit{JWST} observations. We show that the simulated galaxy population is broadly consistent with observations until $z\sim10$. From $z\approx10-12$, the observations indicate a preference for a galaxy population that is largely dust-free, but is still consistent with the simulations. Beyond $z\gtrsim12$, however, our simulation results underpredict the abundance of luminous galaxies and their star-formation rates by almost an order of magnitude. This indicates either an incomplete understanding of the new \jwst data or a need for more sophisticated galaxy formation models that account for additional physical processes such as Population~III stars, variable stellar initial mass functions, or even deviations from the standard \mbox{$\Lambda$CDM} model. We emphasise that any new process invoked to explain this tension should only significantly influence the galaxy population beyond $z\gtrsim10$, while leaving the successful galaxy formation predictions of the fiducial model intact below this redshift. 
\end{abstract}

\begin{keywords}
galaxies: high-redshift, formation -- cosmology: early Universe, first stars -- methods: numerical
\end{keywords}



\section{Introduction}

Until recently, the galaxy population beyond $z\gtrsim8$ was essentially unexplored. \textit{Hubble Space Telescope} (\hst) observations have only been able to detect about $1000$ galaxies at $z=6-8$, and an even lower number at higher redshifts \citep{Livermore2017, Atek2018}. Most of these galaxies do not have a spectroscopic confirmation and only a handful of objects have been observed with complementary facilities such as the Atacama Large millimetre Array \citep[ALMA;][]{Decarli2018, Hashimoto2018} and the \textit{Spitzer Space Telescope} \citep[see for example,][]{Stefanon2021}. However, recent results from \jwst are already revolutionising our study of the processes of galaxy formation and evolution in the early Universe. The large mirror and infrared frequency coverage in principle allows for the detection of rest-frame optical emission of galaxies all the way to $z\sim20$ \citep{Kalirai2018, Williams2018}. In fact, there are already numerous detections of galaxies at $z\gtrsim10$ \citep[see for example,][]{Furtak2022} and and less certain claims of galaxy candidates up to $z\sim17$ \citep{Donnan2022, Harikane2022}. 

Intriguingly, the early science observations with the \jwst show an abundance of high-redshift luminous galaxies \citep{Castellano2022, Donnan2022, Finkelstein2022, Harikane2022, Naidu2022} that are in tension with extrapolated estimates from galaxy formation models that are tuned to match the properties of low-redshift galaxies \citep{Behroozi2019, Behroozi2020}. In fact, some works have found extremely massive galaxies with stellar masses $M_\star \sim 10^{11}\,\Msun$ at $z\gtrsim10$ \citep{Labbe2022, R22}. Given the small survey volume, these galaxies seem to have stellar masses that are larger than the available baryonic reservoir of their host dark matter haloes \citep{BK2022, Lovell2022}. Various explanations for this discrepancy have been proposed, including early dark energy models \citep{EDE2022}, variable stellar initial mass functions \citep[IMF;][]{Steinhardt2022}, higher than expected star-formation conversion efficiencies or a metal-free Population~III stellar population that is extremely UV bright, skewing the mass-to-light ratios of these early galaxies \citep{Inayoshi2022}. However, it is difficult to draw definitive conclusions from these first results as the sources have not yet been spectroscopically confirmed and the photometric candidates are based on reductions using preliminary calibrations. In fact, recent works have pointed out that the high photometric fluxes of these galaxy candidates can be explained by young and low-mass stellar populations, eliminating the need to invoke unrealistically high stellar mass galaxies at these redshifts \citep{Endsley2022}. We also note that there have been a few instances of the same source being detected at different redshifts by different groups \citep{Donnan2022, Zavala2022}. While improvements in the photometric and astrometric calibrations have led to improved redshift estimates for some of the galaxies \citep{Finkelstein2022}, there are still uncertainties due to the sensitivity of the measurement to different galaxy spectral energy distribution (SED) templates used for photometric fitting \citep{Endsley2022} and an inability to properly differentiate between a low-redshift dusty starburst galaxy and a high-redshift source \citep[see for example][]{Naidu2022b}. It is also important to note that these observations cover small fields \mbox{($\lesssim 100$\,arcmin$^2$)} that are subject to significant cosmic variance, especially at these high redshifts \citep{Steinhardt2021}.

To robustly assess the potential tension with theory, it is important to establish reliable baseline predictions from current state-of-the-art galaxy formation models, so that they can be used to compare with these recent and forthcoming observations. While quite a few works have endeavoured to model the galaxy population in the reionization epoch \citep[$6\lesssim z \lesssim10$; ][]{Dayal2014, Gnedin2014, Ni2022}, only a few models have made predictions in the extreme high-redshift regime ($z\geq10$). These include estimates from semi-numerical \citep{Mason2015, Behroozi2020}, semi-analytic \citep{Yung2019} and simulation \citep{Wilkins2022} frameworks. Although hydrodynamical simulations provide a greater understanding of the underlying physical processes that govern the properties of high-redshift galaxies, they are quite expensive which limits their ability to model a large representative volume of the Universe with high-resolution. This is especially important at high redshift, where the source density is so low that only the most massive overdensities have had time to collapse into haloes. This drawback can be overcome by either running large-volume simulations that only simulate the high-redshift Universe \citep{Feng2016, Bird2022} or resimulate sub-regions in a large volume with higher resolution to build up a composite luminosity function at these high redshifts \citep{Lovell2021}.

In this paper, we utilise the new large-volume ($L_\mathrm{box}=738.12\,\mathrm{cMpc}$) hydrodynamical simulation of the MillenniumTNG (\mtng) project to make predictions for the high-redshift ($z\geq8$) galaxy population. This large volume allows us to make predictions for the rare objects that exist at these high redshifts. The simulation uses the IllustrisTNG \citep{Weinberger2017, Pillepich2018b} physics model which has been shown to produce a realistic representation of the galaxy population at low redshifts \citep{Pillepich2018, Nelson2018, Marinacci2018, Tacchella2019}. We supplement the \mtng simulation with three higher-resolution, smaller-volume simulations that share the same underlying galaxy formation model and the \textsc{arepo} code base \citep{Springel2010}, namely \tngs \citep{Pillepich2019}, \thesan\footnote{\thesan uses \textsc{arepo-rt}, which is the radiation hydrodynamic extension to \textsc{arepo}, comprehensively described in \citet{Kannan2019}.} \citep{KannanThesan, GaraldiThesan, SmithThesan}, and \tngl \citep{Springel2018}. Combining these simulations allows us to make predictions over a wide halo mass range \mbox{($8\lesssim \mathrm{log}(M_\mathrm{halo} [\Msun]) \lesssim 12.5$)}. We show simulation predictions for the galaxy stellar mass function, UV luminosity functions (UVLFs), star-formation main sequence, metal content of galaxies, and cosmic star-formation rate density (SFRD). We compare the results to recent observational estimates from \jwst and outline potential reasons for the (mis)match.

This paper is one of several studies introducing the MillenniumTNG (MTNG) project which aside from the hydrodynamic simulation analysed here also includes a series of large dark matter-only simulations, including models with massive neutrinos as an additional hot dark matter component. The companion paper by \citet{CHMTNG} details the full simulation suite, corresponding data products and quantifies basic matter and halo clustering statistics. \citet{PakmorMTNG} focuses on the hydrodynamical full physics simulation of the MTNG project with special emphasis on the properties of galaxy clusters. \citet{BarreraMTNG} presents a novel version of the L-galaxies semi-analytic model of galaxy formation and its application to lightcone outputs of the MTNG simulations. \citet{BoseMTNG} presents a study of galaxy clustering based on colour-selected galaxy samples. \citet{FerlitoMTNG} presents studies of weak gravitational lensing both in the dark matter and full physics runs. \citet{ContrerasMTNG} shows how the cosmological parameters of MTNG can be constrained based on galaxy clustering measurements. \citet{BorMTNGa,BorMTNGb} examines aspects of halo occupation distribution modelling and finally \citet{Delgado2022} studies intrinsic alignments and galaxy shapes. The present work is structured as follows. We introduce our methodology in Section~\ref{sec:methods}, while the main results are presented in Section~\ref{sec:results}. We summarise and discuss our conclusions in Section~\ref{sec:conclusions}.

\begin{table*}
	\centering
	\caption{Main numerical parameters of the different hydrodynamical simulations used in this work. The columns from left to right list the simulation name, box size, total number of particles, mass of dark matter (DM) particles, target mass of gas cells, DM gravitational softening length, and minimum gas softening.}
	\label{tab:simulations}
	\begin{tabular}{lccccccccc} 
		\hline
		Name & $L_\mathrm{box}$ & $N_\mathrm{particles}$ & $m_\mathrm{DM}$ & $m_\mathrm{gas}$ & $\epsilon_\mathrm{DM}$ & $\epsilon^{\mathrm{min}}_\mathrm{gas}$ \\
		& [cMpc] & & [$\Msun$] & [$\Msun$] & [ckpc] & [ckpc] \\
		\hline
		MTNG740 & $738.12$ & $2 \times 4320^3$ & $1.62 \times 10^8$ & $3.10 \times 10^7$ & $3.7$ & $0.37$ \\
		TNG300 & $302.63$ & $2 \times 2500^3$ & $5.90 \times 10^7$ & $1.09 \times 10^7$ & $1.48$ & $0.37$ \\
		THESAN & $95.50$ & $2 \times 2100^3$ & $3.12 \times 10^6$ & $5.82 \times 10^5$ & $2.2$ & $2.2$ \\
		TNG50 & $51.67$ & $2 \times 2160^3$ & $4.58 \times 10^5$ & $8.41 \times 10^4$ & $0.30$ & $0.07$ \\
		\hline
	\end{tabular}
\end{table*}

\section{Methods}
\label{sec:methods}

We use the large-volume full-physics MillenniumTNG (\mtng) hydrodynamic simulation to make predictions for the high-redshift galaxy population. The full simulation suite of MTNG consists of several full physics and dark matter only N-body simulations of various box sizes and resolutions \citep[see][]{CHMTNG}. In this work, we use the largest full-physics simulation of the project, labelled MTNG740, which has a box size of \mbox{$738.12$ cMpc} on a side, resolved by $4320^3$ dark matter and gas particles each, setting the mass resolutions to $1.62 \times 10^8 \, \Msun$ and $3.10 \times 10^7 \, \Msun$ for dark matter and baryons, respectively. The softening length for dark matter and star particles is set to $3.7 \, \mathrm{ckpc}$ while the softening length for the gas varies with the local cell size, constrained by a minimum of $0.37 \, \mathrm{ckpc}$. The simulation is performed using the moving mesh code \textsc{arepo} \citep{Springel2010}, which solves the hydrodynamic (HD) equations on an unstructured Voronoi grid constructed from a set of mesh generating points that are allowed to move along with the underlying gas flow. A quasi-Lagrangian solution to the fluid equations is obtained by solving the Riemann problem at the interfaces between moving mesh cells in the rest-frame of the interface. Gravity is solved with a Tree-PM approach that uses an oct-tree \citep{Barnes1986} algorithm to estimate the short range gravitational forces and a Particle Mesh method \citep{Gadget4} to compute the long range ones. 

\begin{figure*}
	\includegraphics[width=0.99\textwidth]{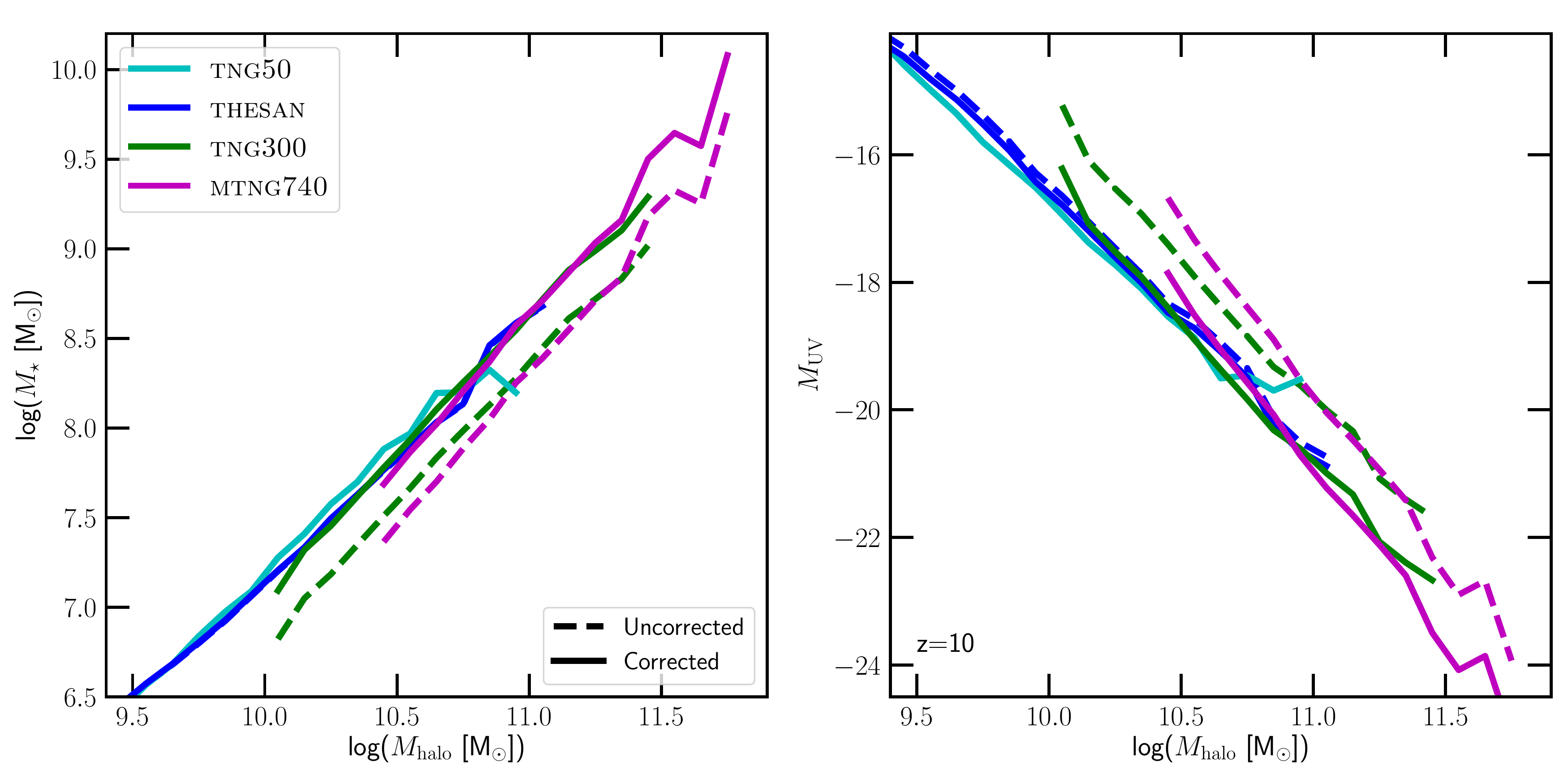}
    \caption{The stellar mass--halo mass (left panel) and UV magnitude--halo mass (right panel) relations for the \tngs (cyan curves), \thesan (blue curves), \tngl (green curves) and \mtng (magenta curves) simulations at $z=10$. The dashed curves show intrinsic predictions while the solid lines are the results after correcting for the different resolutions of the simulations.}
    \label{fig:corr}
\end{figure*}

The simulation uses the IllustrisTNG \citep{Springel2018, Marinacci2018, Naiman2018, Pillepich2018b, Nelson2018, Nelson2019b} galaxy formation model that has empirical prescriptions for processes occurring below the grid scale of the simulation, like star and black hole formation, and associated feedback. The model builds upon the previous Illustris galaxy formation model \citep{Vogelsberger2013, Vogelsberger2014, Vogelsberger2014N} and has improved prescriptions for thermal and kinetic feedback from stars \citep{Pillepich2018} and radio mode feedback from active galactic nuclei \citep[AGN;][]{Weinberger2017}. The interstellar medium (ISM) is modelled as a two-phase gas where cold clumps are embedded in a smooth, hot phase produced by supernova explosions \citep{Springel2003}. Most importantly for our work, this model has been shown to quite successfully reproduce a variety of low-redshift observations, among others, the observed galactic color bimodality \citep{Nelson2018}, color-dependent spatial distribution and clustering of galaxies \citep{Springel2018}, galaxy stellar mass function, galaxy sizes \citep{Genel2018, Pillepich2018b}, metal distributions \citep{Naiman2018}, magnetic field strength and structure \citep{Marinacci2018}, star formation main sequence \citep{Donnari2019} and resolved star formation \citep{Nelson2021}, and galaxy morphologies \citep{RG19, Tacchella2019}. Recently, it has also been used to run coupled galaxy formation and reionization simulations (\thesan) which made predictions for various reionization ($5\lesssim z \lesssim10$) observables like the UV luminosity functions, stellar mass functions and the cosmic star-formation rate density \citep{KannanThesan}. They have also been used to study 21\,cm power spectra \citep{KannanThesan} including an effective bias expansion in redshift space \citep{Qin2022}, IGM--galaxy connections including ionizing mean free path \citep{GaraldiThesan}, Ly$\alpha$ emission and transmission statistics \citep{SmithThesan,Xu2022}, multitracer line intensity mapping \citep{KannanLIM}, and galaxy ionizing escape fractions \citep{Yeh2022}.

\begin{figure*}
	\includegraphics[width=0.99\textwidth]{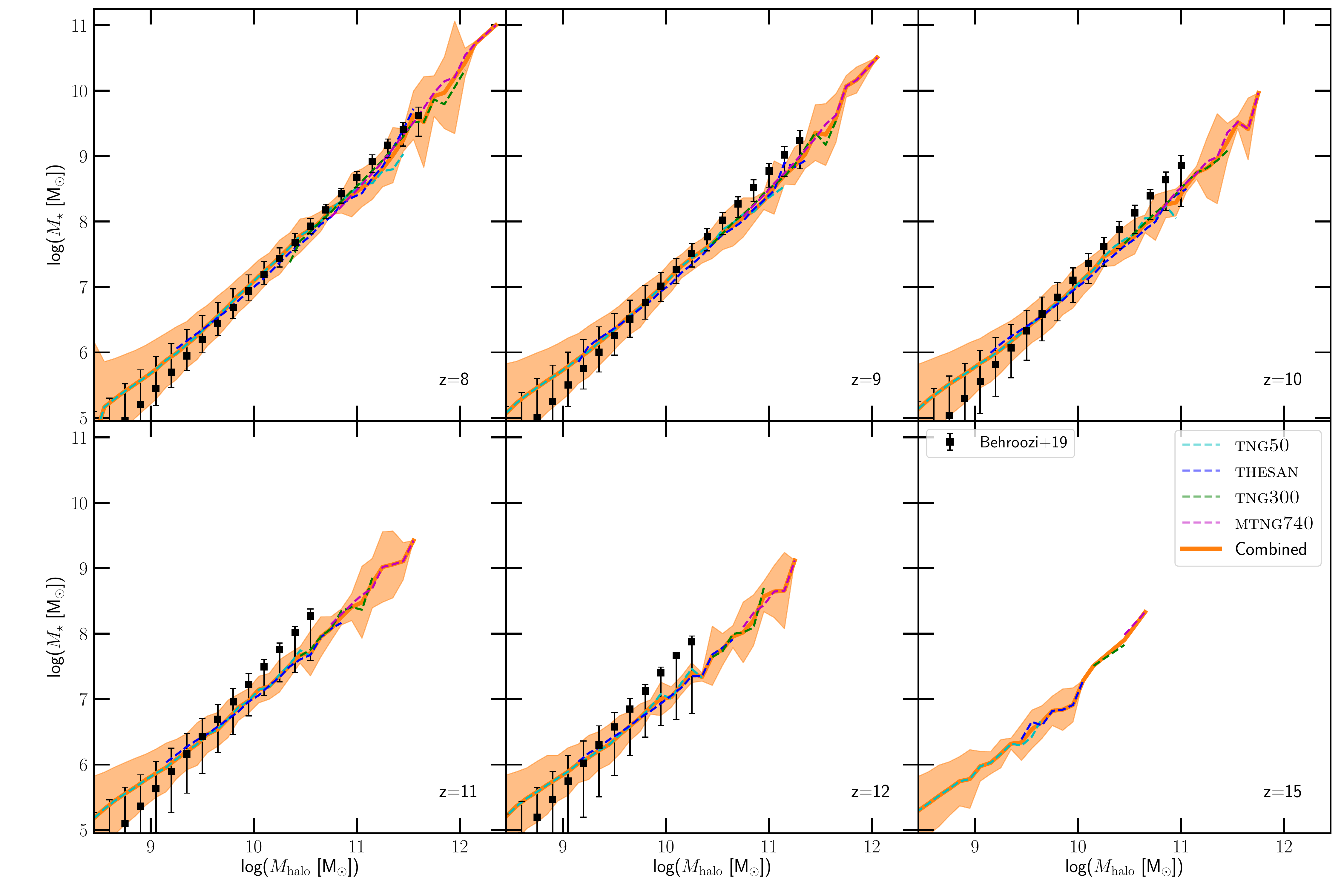}
    \caption{The stellar-halo mass relation at $z=8$ (top left), $z=9$ (top middle), $z=10$ (top right), $z=11$ (bottom left), $z=12$ (bottom middle), and $z=15$ (bottom right). The dashed curves show the relation for \tngs (cyan curves), \thesan (blue curves), \tngl (green curves) and MTNG740 (magenta curves), while the solid orange curves show the combined relation obtained with the methodology described in Section~\ref{sec:methods}. The shaded region denotes the $1\sigma$ dispersion in the relation. The black points are the abundance matching estimates from UNIVERSEMACHINE \citep{Behroozi2019}. The simulations are broadly in agreement with the abundance matching results, but the TNG model seems to predict slightly lower star-formation efficiencies in high-mass haloes ($\mathrm{log}(M_\text{halo}[\Msun]) \gtrsim 10.5$).}
    \label{fig:smhm}
\end{figure*}

\subsection{Combining simulations and resolution corrections}
\label{sec:rescor}
While the MTNG740 hydrodynamical simulation has quite a large volume allowing us to make predictions for the rare objects at high-redshifts, its relatively low resolution does not allow for accurate predictions of low-mass galaxies. To overcome this issue we combine the results of MTNG with three other simulations that use the same underlying galaxy formation model, namely \tngs \citep{Pillepich2019}, \thesan \citep{KannanThesan, GaraldiThesan, SmithThesan}, and \tngl \citep{Springel2018}. These simulations differ from each other in mass and spatial resolution and boxsize, as outlined in Table~\ref{tab:simulations}. The resolution differences lead to moderately different predictions for the galaxy population with the higher-resolution simulations having slightly higher star-formation efficiencies, as shown in \citet{Pillepich2018b}. Therefore, obtaining consistent predictions over a large mass range by combining the various simulations requires correcting for the different resolutions. In this work, we take the predictions of \tngs as the baseline and correct \thesan, \tngl and MTNG740 results to match \tngs. \tngs is used as baseline because it is the highest resolution simulation in our suite. Moreover, it has been shown that for well resolved galaxies, the IllustrisTNG model provides convergent results, with the resolution corrections becoming progressively smaller (or even vanishing) when the resolution is improved \citep{Pillepich2019}. This is seen clearly in the left panel Figure~\ref{fig:corr}, which plots the stellar mass of the galaxy as a function of halo mass. While the low-resolution \tngl \, (dashed green curve) and MTNG740 (dashed magenta curve) simulations have stellar masses that are on average about $\sim 0.3$ dex lower than the \tngs \, (dashed cyan curve) results, the \thesan \, (dashed blue curve) and \tngs \, practically lie on top of each other despite the mass resolutions differing by about a factor of $\approx 7$.

\begin{figure*}
	\includegraphics[width=0.99\textwidth]{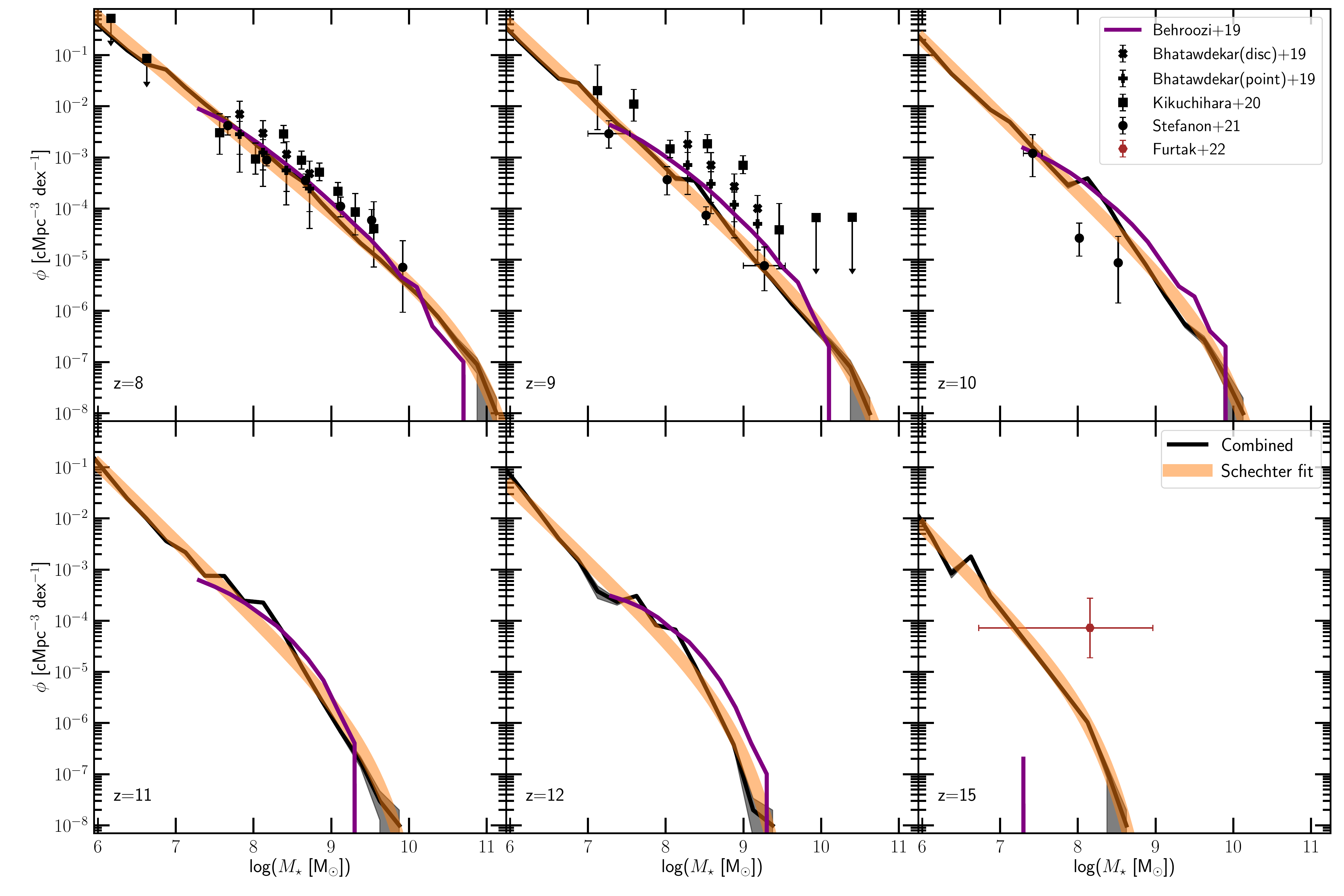}
    \caption{The galaxy stellar mass function at $z=8-15$ as labelled.
    The solid black curve gives the combined stellar mass function, while the orange curve depicts the Schechter function fit (Eq.~\ref{eq:sf_gsmf}; \citealt{SF1976}) for the combined mass function, with the best-fit parameters listed in Table~\ref{tab:sf_gsmf}. For comparison, we show results from \hst and \spitzer observations by \citet[black crosses for disc-like source constraints and black pluses for point-like sources;][]{Bhatawdekar2019}, \citet[black squares;][]{Kikuchihara2020} and \citet[black circles;][]{Stefanon2021}, and recent \jwst results from \citet[brown circles;][]{Furtak2022}.}
    \label{fig:gsmf}
\end{figure*}

We apply a methodology similar to that outlined in \citet{Pillepich2018b} and \citet{Vogelsberger2020} to make these corrections. The impact of the resolution on the halo mass is very small \citep{Jenkins2001}, so it can be used as a stable reference quantity to make the relevant resolution corrections. For each simulation, we divide the galaxies into $50$ logarithmically spaced bins with the halo masses ranging from $10^{7.5}$ to $10^{12.5} \, \Msun$. In each of these bins and for all simulations we calculate the median stellar masses, star-formation rates, stellar and gas metallicities, and UV magnitudes ($M_\mathrm{UV}$) at \mbox{$1500$\,\AA}. For each simulation, we only consider the range with one hundred per cent completeness, i.e., every halo in the particular mass range must contain at least one stellar particle.\footnote{We note that this is different from assuming that every halo with one star particle is resolved. As shown in \citet{Yeh2022}, this resolution constraint is quite stringent and the completeness factor only becomes unity in haloes that are an order of magnitude heavier than the nominal resolution limit of the simulation. This constraint ensures that the median quantities like the stellar-mass halo-mass relation, star forming main sequence, mass metallicity relation etc., are converged (in the sense that they follow the same relation as high mass galaxies if extended to low masses). Since we are mainly interested in the median relations, this resolution constraint works well for the purposes of this paper.} This helps to ensure that only properties of well-resolved haloes are considered in this work. 

Each simulation only covers a portion of the mass range due to the different box sizes and resolutions, with overlapping predictions only available for a relatively small range of masses. To overcome this obstacle the resolution corrections are made in a hierarchical manner; i.e., the corrections for a particular simulation are calculated with respect to the next higher-resolution simulation in the series. So the MTNG740 corrections are calculated by taking \tngl results as the baseline, \tngl uses \thesan, and \thesan uses \tngs. Thus, the total correction for a particular simulation is the sum of the corrections for all higher-resolution simulations. The individual corrections are determined by calculating the average difference between the values of the simulations in the overlapping halo mass range. We note that the average difference is calculated in log space for the stellar masses, SFRs and metallicities ($Z$), while the UV magnitude corrections are calculated as is, because these are already proportional to the log of luminosity. The correction is then applied to all galaxies in the simulation, irrespective of whether they lie in the overlap region or not. This amounts to adding a constant calibration offset to all galaxies from lower-resolution simulations. 

In Figure~\ref{fig:corr}, we show the outcome of this correction process on the predicted stellar mass--halo mass (left panel) and $M_\mathrm{UV}$--halo mass (right panel) relations. The dashed curves show the uncorrected values for \tngs (cyan curves), \thesan (blue curves), \tngl (green curves), and MTNG740 (magenta curves), respectively, at $z=10$, while the corresponding solid curves denote the corrected values. This approach of correcting for resolution in a hierarchical manner is well-suited for producing consistent predictions for a variety of quantities over a wide mass range. Finally, the corrected values of the different simulations are combined together to make a single forecast for the entire galaxy population. The stellar mass/luminosity functions are combined using $\Phi_\mathrm{combined} = \Sigma (\Phi_i n_i^2)/(\Sigma n_i^2)$ where $i \in \{\tngs, \thesan, \tngl, \textsc{MTNG740}\}, \Phi$ is the mass/luminosity function, and $n$ is the number of galaxies in the corresponding mass/luminosity bin \citep{Vogelsberger2020}. Other quantities $Q$ (like the mean SFR) that do not rely on the number of galaxies within a certain bin are combined by using $Q_\mathrm{combined} = \Sigma(Q_i n_i)/\Sigma(n_i)$.

\section{Results}
\label{sec:results}
In Figure~\ref{fig:smhm} we plot the stellar mass--halo mass (SMHM) relation at $z=8$ (top left), $z=9$ (top middle), $z=10$ (top right), $z=11$ (bottom left), $z=12$ (bottom middle), and $z=15$ (bottom right). The dashed curves show the resolution-corrected estimates for \tngs (cyan curves), \thesan (blue curves), \tngl (green curves) and MTNG740 (magenta curves), while the solid orange curves and corresponding shaded regions give the median and $1\sigma$ distribution of the simulated SMHM relation derived by combining all simulations using the method described in Section~\ref{sec:rescor}. For comparison, we also include results from the abundance matching estimates from UNIVERSEMACHINE \citep[black points; ][]{Behroozi2019}. Our simulation results are in excellent agreement with these independent predictions, although the slopes of the relation are slightly different. The TNG model predicts slightly higher (lower) stellar masses at low (high) halo masses, with the discrepancy growing with increasing redshift. This provides further confidence that the procedures for resolution corrections and combining the different simulations used in this work are capable of providing reliable predictions over such a wide halo mass range.

\begin{table}
	\centering
	\caption{Parameters for Schechter function fits of the galaxy stellar mass function at different redshifts (see equation~\ref{eq:sf_gsmf}).}
	\label{tab:sf_gsmf}
	\begin{tabular}{cccc} 
		\hline
		$z$ & $\alpha$ & $\mathrm{log}(M^*)$ & $\mathrm{log}(\Phi^*)$ \\
		& & [log($\Msun$)] & [cMpc$^{-3}$ dex$^{-1}$] \\
		\hline
		8 & $-2.28 \pm 0.02$ & $10.68 \pm 0.04$ & $-6.60 \pm 0.10$ \\
		9 & $-2.42 \pm 0.03$ & $10.25 \pm 0.08$ & $-6.83 \pm 0.19$ \\
		10 & $-2.49 \pm 0.06$ & $9.66 \pm 0.09$ & $-6.51 \pm 0.24$ \\
        11 & $-2.49 \pm 0.08$ & $9.32 \pm 0.10$ & $-6.26 \pm 0.29$ \\
        12 & $-2.48 \pm 0.13$ & $8.71 \pm 0.10$ & $-5.74 \pm 0.34$ \\
        15 & $-2.56 \pm 0.19$ & $8.01 \pm 0.12$ & $-5.66 \pm 0.43$ \\
		\hline
	\end{tabular}
\end{table}

Next, we turn our attention to Fig.~\ref{fig:gsmf}, which plots the galaxy stellar mass function $\Phi$ (in units of $\mathrm{cMpc}^{-3} \, \mathrm{dex}^{-1}$) at $z=8-15$, as indicated. The black solid curves show the simulation estimates while the orange curves are Schechter function fits \citep{SF1976}, given by 
\begin{equation}
    \Phi\,\, {\rm d} \, \mathrm{log}M = \mathrm{ln}(10) \,\Phi^* \frac{\left[10^{(\mathrm{log}M - \mathrm{log}M^*)}\right]^{\alpha+1}}{e^{10^{(\mathrm{log}M - \mathrm{log}M^*)}}} \,{\rm d} \, \mathrm{log}M \, ,
    \label{eq:sf_gsmf}
\end{equation}
where $M$ is the stellar mass of the galaxy (in units of $\Msun$), $\alpha$ is the low-mass slope, $M^*$ is the stellar mass above which the mass function cuts off exponentially, and $\Phi^*$ is the value of the mass function at $M^*$. The values of these fits are listed in Table~\ref{tab:sf_gsmf}. The plot also shows observational estimates using lensed galaxy observations in the Hubble Frontier Fields from \citet[black crosses for disc-like source constraints and black pluses for point-like sources; ][]{Bhatawdekar2019} and \citet[black squares; ][]{Kikuchihara2020}, \spitzer/IRAC measurements by \citet[black circles; ][]{Stefanon2021}\footnote{The stellar mass estimates from \citet{Stefanon2021} have been converted to match the Chabrier IMF \citep{Chabrier2003} used in this work, reducing them by a factor of $1.7$.} and the recent gravitationally lensed galaxy candidates detected behind the galaxy cluster SMACS J0723.3-7327 \citep[brown circle; ][]{Furtak2022}. The previous \hst/\spitzer estimates are in general agreement with the simulation results up to $z\sim10$ over a wide mass range. On the other hand, the new $z\sim15$ \jwst observations seem to indicate an overabundance of massive galaxies in the early Universe.

\begin{figure*}
	\includegraphics[width=0.99\textwidth]{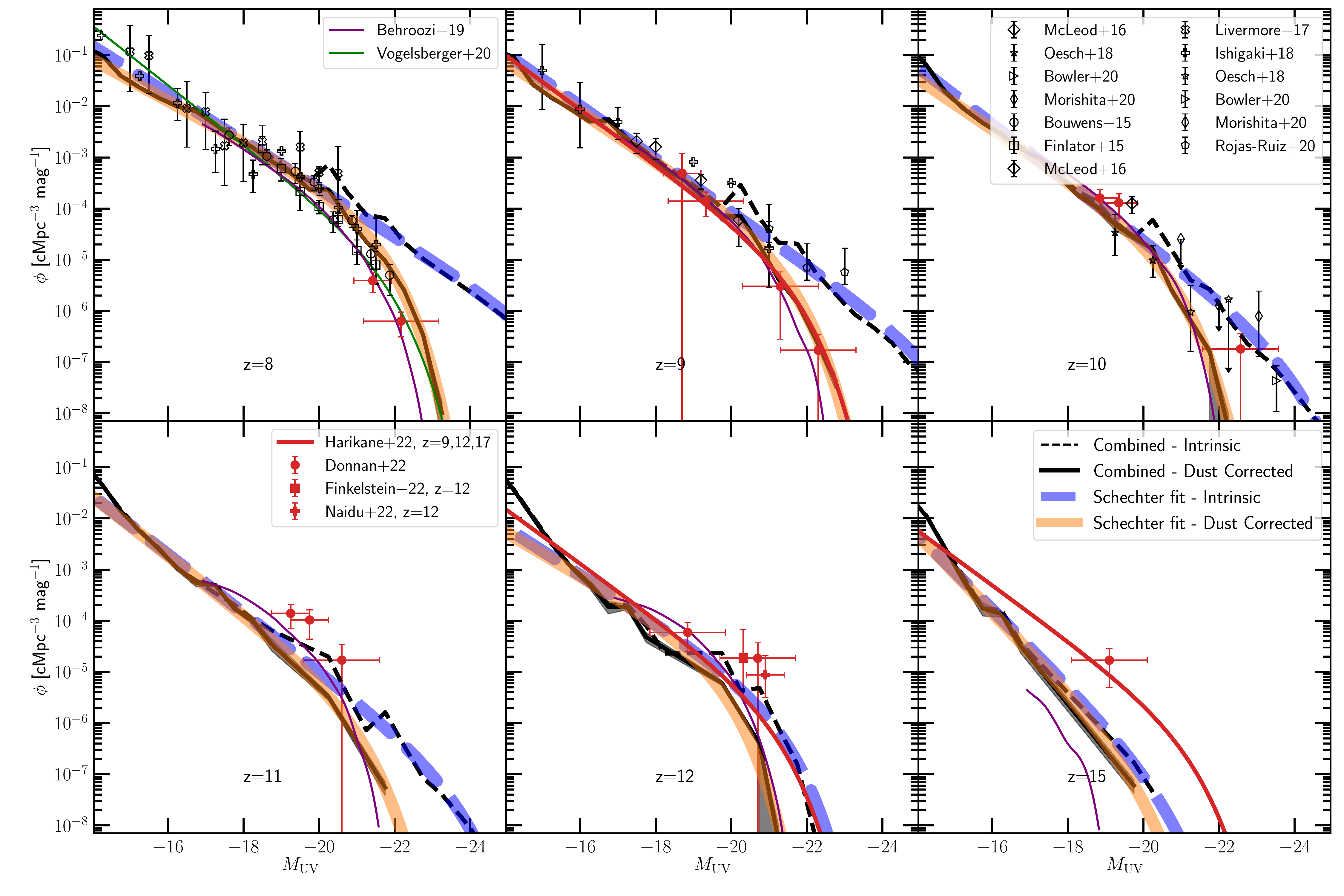}
    \caption{The UV luminosity function at $z=8-15$ as labelled. The dashed black curves give the intrinsic UV luminosity function while the solid black curves show the dust-corrected values. The Schechter function fits for the intrinsic and dust-corrected UVLFs are shown as dashed blue and solid orange curves, respectively. For comparison, abundance matching estimates from \citet{Behroozi2019} are shown in purple, while the Schechter function fit to the estimates from the post-processing dust radiative transfer calculations by \citet{Vogelsberger2020} are shown in green. Pre-\jwst measurements from \citet{Bouwens2015}, \citet{Finkelstein2015}, \citet{McLeod2016}, \citet{Livermore2017}, \citet{Ishigaki2018}, \citet{Oesch2018}, \citet{Bowler2020}, \citet{Mor2020}, and \citet{RR2020} are shown with black symbols as indicated. The new measurements from \jwst, from \citet{Donnan2022}, \citet{Finkelstein2022}, and \citet{Naidu2022}, are plotted as filled red symbols, while the Schechter function fits to the observational estimates from \citet{Harikane2022} are drawn as red curves.}
    \label{fig:uvlf}
\end{figure*}

A more direct comparison with observations can be made by considering the UV luminosity function at $z=8-15$, as shown in Figure~\ref{fig:uvlf}. For the \tngs, \thesan, and \tngl simulations this is obtained by summing up the radiation output at rest-frame \mbox{$1500$ \AA} (using BPASS version 2.2.1 tables; \citealt{BPASS2017}) of all the stars in the identified subhalo.\footnote{In this work we have not included nebular emission lines, which can in principle contribute significantly to the photometric fluxes, especially in high-redshift, low-metallicity environments. They have the ability to bias photometric estimates of the stellar mass and star formation rates \citep{Endsley2022}. However, there are very few strong emission lines around \mbox{$1500$\,\AA}, so our theoretical estimates of $M_\mathrm{UV}$ should be quite robust.} We note that due to the probabilistic nature of the star-formation routine, the star-formation history will only be sparsely sampled in haloes with low SFR. These haloes will have long periods with zero star formation interspersed with sudden jumps in SFR as a new particle is stochastically spawned. This young and massive star will then dominate the entire radiation output of the galaxy, especially if the mass of the galaxy is close to the resolution limit of the simulation. This will adversely affect the UV luminosity function of the simulation. To overcome this numerical artefact, the age and mass of stars formed less than \mbox{$5$\,Myr} ago are smoothed over a timescale given by $t_\mathrm{smooth} = \Sigma M_\star (< 5 \mathrm{Myr}) / \mathrm{SFR_{gal}}$, where $\mathrm{SFR_{gal}}$ is the instantaneous SFR of the corresponding galaxy calculated by summing up all the SF probabilities of the cells on the EoS \citep[see][for more details]{Springel2003}. This smoothing procedure is only done for haloes with \mbox{$t_\mathrm{smooth}> 5$ Myr}. We note that this only affects haloes close to the resolution limit and allows for a more faithful prediction of the simulated UV luminosity function \citep{KannanThesan}.

This method requires a knowledge of the ages and metallicities of all stars in the halo. While this information is available in the full simulation snapshots, the MTNG740 simulation only outputs the SUBFIND and friends-of-friends halo catalogues at $z\gtrsim7$. Therefore, it is not possible to derive the UV magnitudes at \mbox{$1500$\,\AA} using raw particle/cell data. However, the group catalogue output includes the magnitudes in eight bands (U, B, V, K, g, r, i, z) based on the summed luminosities of all the stellar particles in each group \citep{Torrey2014}. These eight magnitudes are also derived from the distribution of stellar ages and metallicities, and therefore contain all the information needed to predict the value at \mbox{$1500$ \AA} as well. We exploit this by training a Ridge regression model on the eight band outputs of the group catalogues to predict the \mbox{$1500$\,\AA} magnitude using values from the \tngs, \thesan, and \tngl simulations. This model is then employed to forecast $M_\mathrm{UV}$ at \mbox{$1500$\,\AA} for galaxies in the MTNG740 simulation.

\begin{table}
	\centering
	\caption{Schechter function fits for the galaxy luminosity functions.}
	\label{tab:sf_uvlf}
	\begin{tabular}{cccccccccc} 
		\hline
		$z$ & $\alpha_\mathrm{UV}$ & $M^*_\mathrm{UV}$ & $\mathrm{log}(\Phi^*_\mathrm{UV})$ \\
		& & [mag] & [cMpc$^{-3}$ mag$^{-1}$] \\
		\hline
		8 & $-2.01 \pm 0.03$ & $-21.11 \pm 0.04$ & $-3.92 \pm 0.06$ \\
        9 & $-2.23 \pm 0.05$ & $-21.06 \pm 0.10$ & $-4.54 \pm 0.12$ \\
        10 & $-2.25 \pm 0.05$ & $-20.29 \pm 0.07$ & $-4.54 \pm 0.11$ \\
        11 & $-2.44 \pm 0.06$ & $-20.62 \pm 0.14$ & $-5.35 \pm 0.17$ \\
        12 & $-2.06 \pm 0.27$ & $-19.28 \pm 0.27$ & $-4.47 \pm 0.41$ \\
        15 & $-2.91 \pm 0.39$ & $-19.50 \pm 1.26$ & $-6.45 \pm 1.52$ \\
		\hline
		 & & Intrinsic & \\
		\hline
		8 & $-2.16 \pm 0.04$ & $-26.07 \pm 0.19$ & $-6.40 \pm 0.17$ \\
        9 & $-2.30 \pm 0.06$ & $-25.36 \pm 0.47$ & $-6.83 \pm 0.38$ \\
        10 & $-2.42 \pm 0.05$ & $-23.69 \pm 0.23$ & $-6.68 \pm 0.23$ \\
        11 & $-2.43 \pm 0.06$ & $-23.34 \pm 0.29$ & $-6.87 \pm 0.29$ \\
        12 & $-2.18 \pm 0.16$ & $-20.90 \pm 0.24$ & $-5.39 \pm 0.33$ \\
        15 & $-2.84 \pm 0.28$ & $-20.28 \pm 1.46$ & $-6.89 \pm 1.57$ \\
        \hline
	\end{tabular}
\end{table}

Attenuation by dust grains is especially important for the high-luminosity end. However, the amount of dust and its composition is not well constrained at these high redshifts. Therefore, we use an empirical dust-attenuation ($A_\mathrm{UV}$) model, which is obtained by fitting the IRX--UV relationship inferred from ALMA observations at $z\sim4-7$
in \citet{Bouwens2016} to the following equation:
\begin{equation} \label{eq:d1}
    A_\mathrm{UV} = 2.5 \, \log\left(1+10^{0.4\alpha_\mathrm{dust}(M_\mathrm{dust}(z) -M_\mathrm{UV, int})}\right) \, ,
\end{equation}
where
\begin{equation} \label{eq:d2}
    M_\mathrm{dust}(z) = M_\mathrm{dust,0} + z \cdot M_{\mathrm{dust},z} \, ,
\end{equation}
and the observed UV magnitude ($M_\mathrm{UV}$) is $M_\mathrm{UV, int} + A_\mathrm{UV}$ \citep{Behroozi2020}. $M_{\mathrm{dust},z}$ is the magnitude below which the dust attenuation rises exponentially and it is a linear function of redshift (z) with $M_\mathrm{dust,0}$ and $M_{\mathrm{dust},z}$ being the intercept and slope of the relation, respectively. Finally, $\alpha_\mathrm{dust}$ is an opacity parameter that is roughly proportional to the optical depth at \mbox{$1500$\,\AA}. We plot both the intrinsic (dashed black curves) and dust-attenuated (solid black curves) UV luminosity function, with the corresponding Schechter function fits for the intrinsic (dashed blue curves) and dust-attenuated UVLFs indicated by dashed blue and solid orange curves, respectively. The corresponding best-fit parameters are given in Table~\ref{tab:sf_uvlf}. For comparison, we also plot the semi-empirical estimates from \citet[purple curves;][]{Behroozi2019} and post-processing dust radiative transfer calculations of TNG galaxies outlined in \citet[green curve;][]{Vogelsberger2020}. The results from \citet{Behroozi2019} are generally in good agreement with the dust-attenuated UVLFs, with only slight differences at $z=15$. The \citet{Vogelsberger2020} results generally agree with ours, because we are looking at mostly the same galaxy population, only extended to higher luminosities due to the addition of the MTNG740 results. However, slight differences arise from the fact that the model for generating galactic SEDs is different between the two works. The pre-\jwst estimates of the UVLF from \citet{Bouwens2015}, \citet{Finkelstein2015}, \citet{McLeod2016}, \citet{Livermore2017}, \citet{Ishigaki2018}, \citet{Oesch2018}, \citet{Bowler2020}, \citet{Mor2020}, and \citet{RR2020} are presented as black unfilled symbols, as indicated. The recent \jwst observational estimates from \citet{Donnan2022}, \citet{Finkelstein2022}, and \citet{Naidu2022} are depicted as red filled symbols, while the \citet{Harikane2022} Schechter function fits at $z=9$, $12$, and $17$ are shown as red solid curves. 

\begin{figure*}
	\includegraphics[width=0.99\textwidth]{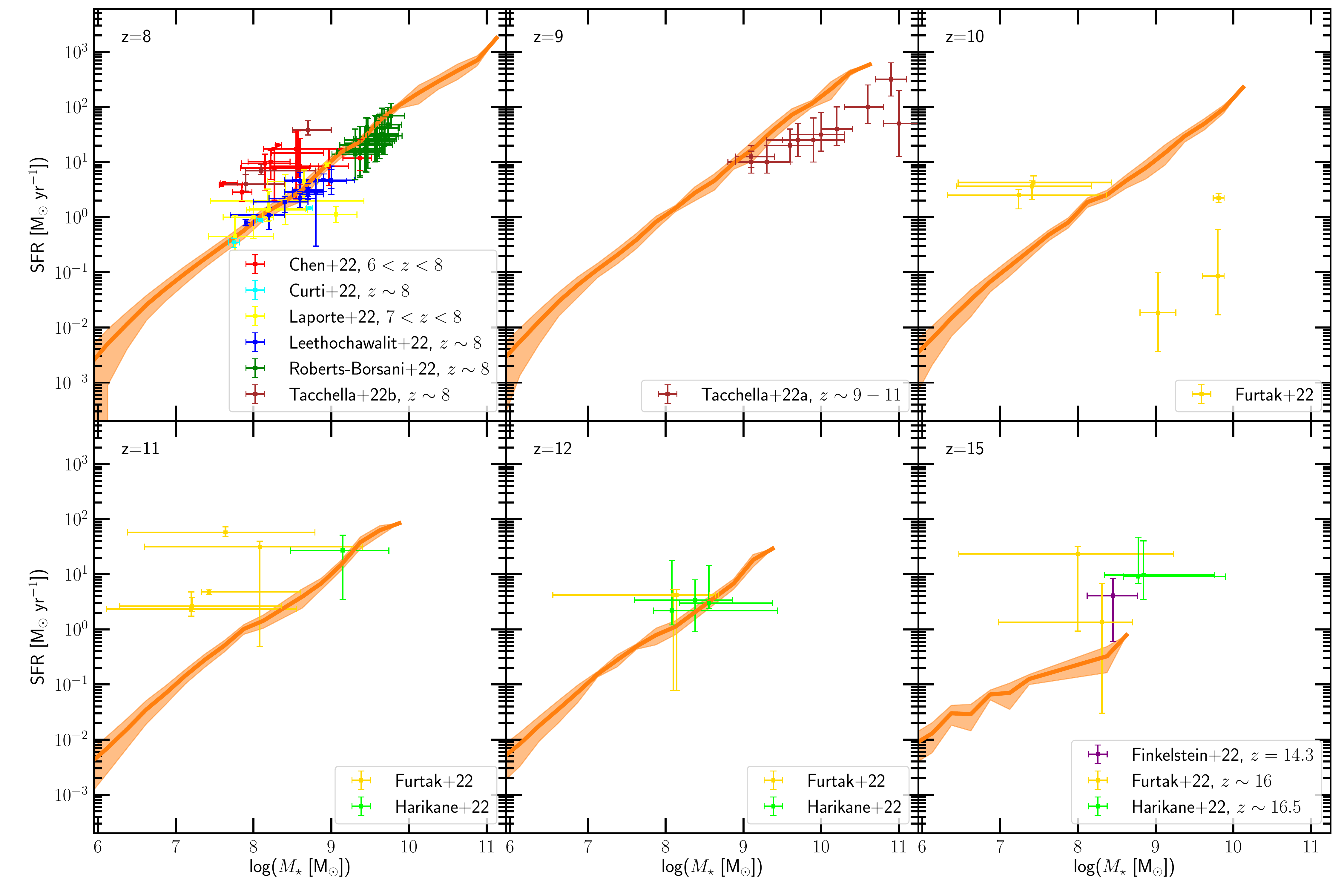}
    \caption{The predicted star-formation rate as a function of the stellar mass at $z=8-15$ as labelled. The simulation results and their $1\sigma$ distributions are indicated by the orange curves and shaded regions, respectively. Observational estimates from \citet[red points;][]{Chen2022}, \citet[cyan points;][]{Curti2022}, \citet[purple points;][]{Finkelstein2022}, \citet[gold points;][]{Furtak2022}, \citet[light green points;][]{Harikane2022}, \citet[yellow points;][]{Laporte2022}, \citet[blue points;][]{Lee2022}, \citet[dark green points;][]{RB2022}, and \citet[maroon points][]{Tacchella2022b, Tacchella2022a} are also shown for comparison.}
    \label{fig:sfr}
\end{figure*}

The dust-attenuated luminosity functions faithfully reproduce the observational estimates from both \hst and \jwst measurements at $z=8$ and $9$. At $z=10$, however, there seems to be a higher than theoretically expected abundance of luminous ($M_\mathrm{UV}<-22$) galaxies. In fact, the observations seem to prefer the dust-free UVLF, suggesting that most of the galaxies, even the brightest ones, are mainly dust free at these redshifts. At $z=11$ and $z=12$, the new \jwst measurements seem to be slightly higher but still consistent with the simulated dust-free UVLF. This conclusion is supported by recent targeted ALMA observations of GHZ2/GLASS-z13, one of the brightest and most robust candidates at $z>10$, by \citet{Bakx2022} who were unable to detect any dust continuum, indicating negligible dust content in these high-redshift galaxies. However, by $z=15$, the observed abundance of galaxies is about an order of magnitude higher than the simulated predictions.

These results seem to suggest that the star-formation and stellar feedback routines used in the TNG model, when combined with the adopted dust-attenuation estimates, do a good job of reproducing the galaxy population for $z\le10$. For $z=10-12$, the observations seem to prefer a dust-free galaxy population \citep{Ferrara2022}, which is still consistent with the galaxy formation models that have been calibrated to match local Universe observations. However, the $z=15$ discrepancy seems to suggest, at least when taken at face value, that we need to reexamine high-redshift galaxy formation physics and/or underlying cosmology. We will discuss this further in Section~\ref{sec:conclusions}. 

\begin{figure*}
	\includegraphics[width=0.99\textwidth]{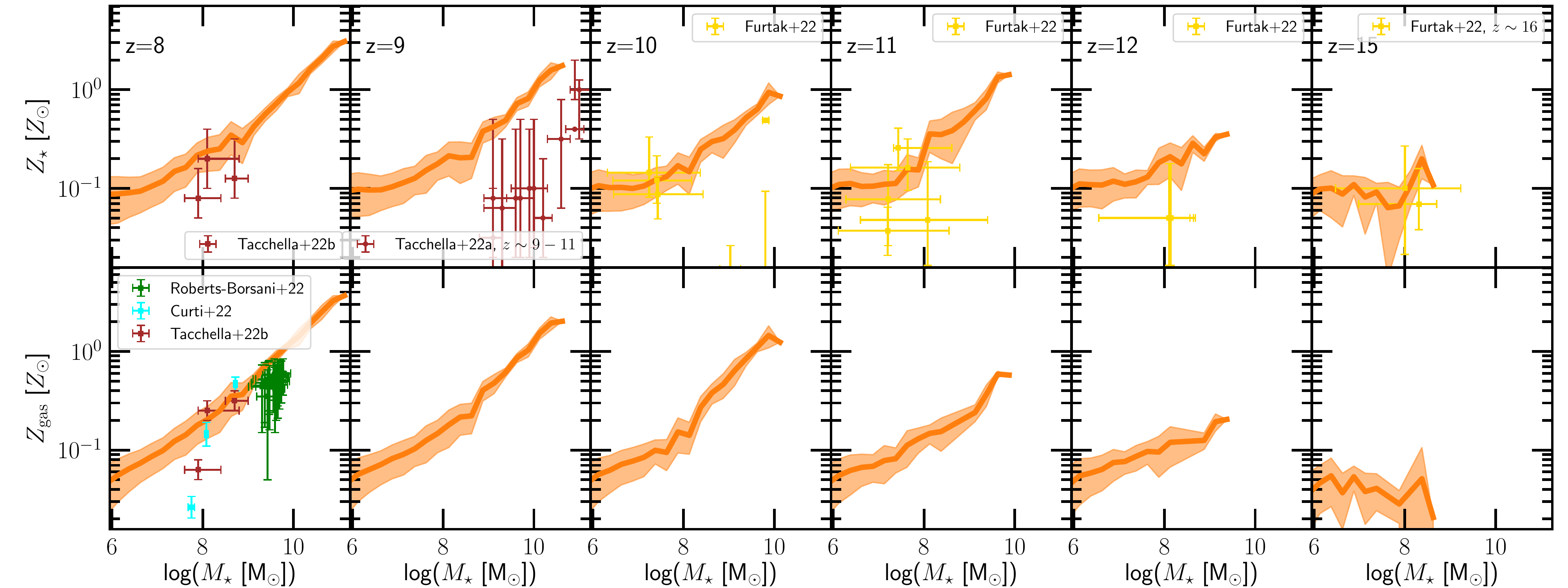}
    \caption{The predicted stellar (top rows) and gas phase (bottom rows) metallicity as a function of the stellar mass at $z=8-15$ as labelled. The plot also includes observational estimates from \citet[cyan points;][]{Curti2022}, \citet[gold points;][]{Furtak2022}, \citet[dark green points;][]{RB2022} and \citet[maroon points][]{Tacchella2022b, Tacchella2022a}.}
    \label{fig:met}
\end{figure*}

Additional insights into the reasons for this discrepancy can be inferred by inspecting the star-formation rate--stellar mass relation, also known as the star-formation main sequence \citep{Whitaker2014}. While the UVLF and GSMF can be biased by factors like cosmic variance \citep{Steinhardt2021}, the star-formation rate as a function of stellar mass is largely independent of the environment of galaxies and is, therefore, an excellent probe for the efficiency with which galaxies form stars. Figure~\ref{fig:sfr} shows the simulated (orange curves) star-formation main sequence at $z=8-15$. A collection of observational estimates from a combination of \hst, \spitzer/IRAC, and \jwst observations presented in \citet{Chen2022}, \citet{Curti2022}, \citet{Finkelstein2022}, \citet{Furtak2022}, \citet{Harikane2022}, \citet{Laporte2022}, \citet{Lee2022}, \citet{RB2022} and \citet{Tacchella2022b, Tacchella2022a} is shown as indicated. There is generally good agreement between the simulations and observational points, at least up to $z\simeq12$, even with the recent \jwst results. Only the SFRs of the gravitationally lensed galaxies detected behind the galaxy cluster SMACS J0723.3-7327 from \citet{Furtak2022} seem inconsistent. They estimate SFRs that are about $1-2$ orders of magnitude above the simulated main sequence at the low-mass end ($M_\star<10^8\,\Msun$), while the SFRs of high-mass galaxies ($M_\star>10^9\,\Msun$) are below the main sequence by a similar amount. However, we do note that the stellar mass measurements have very large uncertainties due to the complexities of creating an accurate lensing model \citep[see for example][]{Atek2018}. Above $z>12$ however, almost all the observational measurements predict systematically higher SFRs by about a factor of $\sim10$, indicating missing physical processes that are not currently modelled in most simulation frameworks.

In a similar vein, the metal content of both the stellar and gaseous components of galaxies helps to constrain important physical processes that govern star formation and feedback, e.g., the amount of fuel available for star formation, where stars are formed, how metal enrichment by stars proceeds, and the role that outflows play in ejecting both mass and metals from galaxies \citep{Tremonti2004}. Stellar metallicity also influences the UV luminosity of stars, with low-metallicity stars emitting a larger number of high-energy photons \citep{BPASS2017}. Moreover, increased metallicity can also lead to larger radiative cooling rates, which in turn provide more fuel for star formation \citep{Wiersma2009}. The average metallicity within galaxies is known to have a tight correlation with stellar mass, with low-mass galaxies being less metal enriched than the high-mass ones \citep{Kewley2008}.

In Figure~\ref{fig:met} we therefore plot the mass-weighted stellar (top panels) and gas phase (bottom panels) metallicity (scaled to the solar metallicity value $\Zsun=0.0127$) of galaxies as a function of their stellar mass. As expected, there is generally a positive correlation between the metallicity and stellar mass, however, the slope of this relation becomes shallower with increasing redshift. We note that this is in disagreement with previous work which found a very weak evolution of the slope of the mass--metallicity relation in the TNG model at $z=2-10$ \citep{Torrey2019}. This might be due to the fact that these early galaxies are so young that there has not been enough time to establish a mass--metallicity relation. Even a single enrichment event can potentially increase the metallicity of low-mass galaxies to the levels seen in the high-mass ones. As expected the stellar metallicities are higher than gas phase ones, because the stars preferentially form in high-density gas that tends to be found in the metal enriched centres of galaxies. Interestingly, the observational estimates from \citet{Curti2022}, \citet{Furtak2022}, \citet{RB2022}, and \citet{Tacchella2022a, Tacchella2022b}, mainly derived from SED modelling of NIRCam photometry, are in good agreement with the simulated results even at $z=15$. The fact that even the low-mass galaxies with very high SFRs, presented in \citet{Furtak2022}, lie on the simulated mass--metallicity relation seems to imply that these galaxies are very young and are undergoing a massive starburst. However, we emphasize the high-mass galaxies that have very low SFRs are consistent with having hardly any metal enrichment at all ($<0.01\,\Zsun$). From a theory standpoint it is quite difficult to explain how these galaxies got quite so massive without enriching their gaseous and stellar components.

\begin{figure}
	\includegraphics[width=0.99\columnwidth]{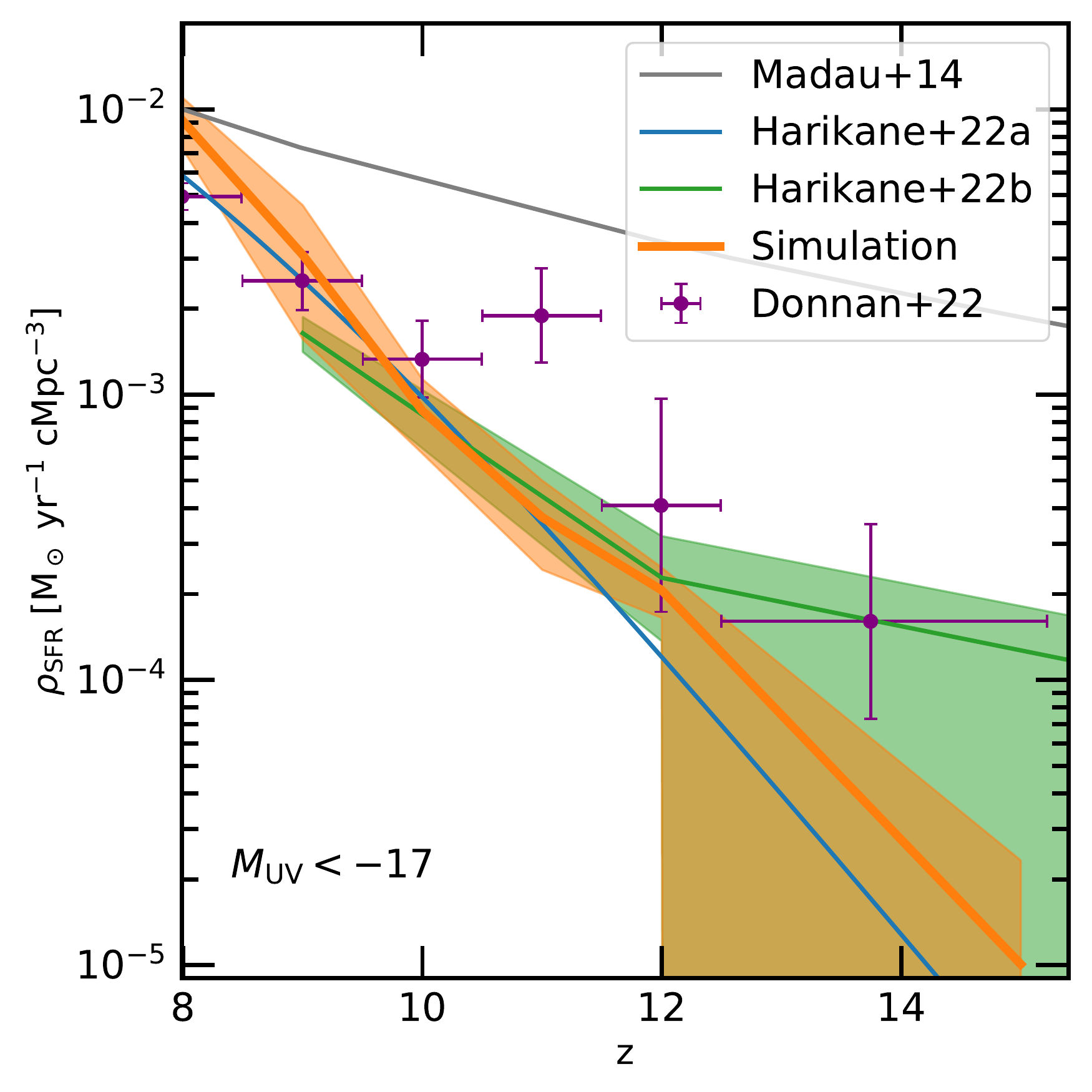}
    \caption{The cosmic star-formation rate density (SFRD) for galaxies with $M_\mathrm{UV}<-17$. The simulation estimates (orange curve and shaded region) are calculated by using the Schechter function fits to the dust-free UVLF, and assuming a constant UV luminosity to star-formation rate ratio given by Eq.~(\ref{eq:sfruv}). The extrapolation from low-redshift fits from \citet{Madau2014} is shown as the grey line. The green curve gives the estimate from \citet{Harikane2022} using the Schechter function fits outlined in their paper. Finally, the purple points display the estimates from \citet{Donnan2022}.}
    \label{fig:sfrd}
\end{figure}

Finally, we turn our attention to the cosmic star-formation rate density as a function of redshift (Figure~\ref{fig:sfrd}), which is derived by integrating the Schechter function fits of the dust-free galaxy luminosity function (Table~\ref{tab:sf_uvlf}) down to $M_\mathrm{uv}=-17$, and converting the UV luminosity to a SFR using the relation
\begin{equation} \label{eq:sfruv}
    \mathrm{SFR \, [\Msun \, yr^{-1}]} = \kappa_\mathrm{UV} L_\mathrm{UV} \, \mathrm{[erg \, s^{-1} \, Hz^{-1}]} \, ,
\end{equation}
where $\kappa_\mathrm{UV}$ is a conversion factor that depends on the recent star-formation history, metal enrichment history, and initial mass function. To be consistent with the observational estimates we use $\kappa_\mathrm{UV} = 1.15 \times 10^{-28} \, \mathrm{\Msun \, yr^{-1}/(erg \, s^{-1} \, Hz^{-1}})$, which is valid for a \citet{Salpeter1995} IMF and consistent with the cosmic star-formation history out to $z\sim8$ \citep{Madau2014}. The plot also shows the SFR density estimates from the best-fit function at $z\lesssim8$ determined by \citet[grey curve;][]{Madau2014} extrapolated beyond $z>8$, and the estimates from \citet{Harikane2022} which were derived assuming a constant star-formation efficiency. The newly derived observational estimates from \jwst observations reported in \citet[purple points;][]{Donnan2022} and \citet[green curve;][]{Harikane22a}\footnote{We note that the cosmic star-formation rate density attributed to \citet{Harikane22a} is derived by using the Schechter function fits quoted in their paper, instead of the double power-law fits, in order to be consistent with the framework used in our current work.} for the very first time put constraints on the star-formation rate beyond $z>10$. The larger number of observed luminous galaxies at $z\gtrsim13$ and the generally higher star-formation rates lead to a cosmic star-formation rate density that is about a factor of $2-5$ higher than the simulation results. However, the simulation predictions are fairly consistent with the \jwst observations at lower redshifts ($z\lesssim12$). We therefore conclude that the simulations produce a realistic galaxy population below $z\leq12$ but there might be some important physical processes missing at even higher redshifts. 

\section{Discussion and Conclusions}
\label{sec:conclusions}

In this work we have presented a variety of predictions for the high-redshift ($z\geq8$) galaxy population from the new MillenniumTNG (MTNG740) hydrodynamical simulation. In order to obtain consistent predictions over a wide halo \mbox{($8\lesssim \mathrm{log}(M_\mathrm{halo} [\Msun]) \lesssim 12.5$)} and stellar \mbox{($5\lesssim \mathrm{log}(M_\star [\Msun]) \lesssim 11$)} mass range, we combine the results of MTNG740 with \tngs \citep{Pillepich2019, Nelson2019}, \thesan \citep{KannanThesan, GaraldiThesan, SmithThesan}, and \tngl \citep{Springel2018, Marinacci2018, Naiman2018, Nelson2018, Pillepich2018b}, which all use the same underlying galaxy formation model. The different resolutions of these simulations lead to moderately offset predictions for the galaxy population, with the higher-resolution simulations having slightly higher star-formation efficiencies. We correct for this using the calibration methodology outlined in Section~\ref{sec:rescor}, using \tngs (the highest resolution simulation in this suite) as baseline.

We make predictions for a variety of high-redshift galaxy properties, including the stellar-halo mass relation, galaxy stellar mass function, UV luminosity function, star-formation main sequence, metal enrichment in both stellar and gas phases, and cosmic star-formation rate density. These are compared to the recent observational estimates from \jwst to determine the validity of the galaxy formation model (which has been calibrated to match the galaxy population in the local Universe) in this high-redshift frontier. The main results and conclusions of this work are as follows:

\begin{enumerate}
    \item The stellar-halo mass relation matches estimates from the abundance matching results of the {\small UNIVERSEMACHINE} model \citep{Behroozi2019}.
    
    \item The predicted galaxy stellar mass functions match the observed estimates up to $z\sim10$, but are below the 
    new \jwst estimates at $z=15$.
    
    \item The simulated UV luminosity function is consistent with both the \hst and \jwst estimates at $z=8$ and $9$. At $z=10$, the observations seem to prefer a galaxy population that is largely free of dust. The $z=11$ and $12$ dust-free luminosity functions are slightly lower but still consistent with the new \jwst measurements. However, by $z=15$, the observed abundance of luminous galaxies is about an order of magnitude higher than the simulated results. Similar results have also been found by other galaxy formation models that have been successful in reproducing the properties of low-redshift galaxies \citep[see for example, ][]{Moster2018, Yung2019, Behroozi2020, Wilkins2022} although the degree to which they disagree with the new observations is different across various models.
    
    \item This behaviour is reflected in the star-formation main sequence and the cosmic star-formation rate density (for galaxies with $M_\mathrm{UV}<-17$), with the simulated results generally being consistent with observational estimates below $z\lesssim12$, but being lower by about an order of magnitude at $z=15$.
    
    \item The metal content in galaxies seems to be consistent with the observations, despite some of the observations showing higher than expected star-formation rates. This suggests that most of the observationally detected galaxies have very young ages with most of the stars in these galaxies being formed very recently \citep{Furtak2022, Mason2022}.
    
    \item The abundance of massive luminous galaxies observed with \jwst seems to suggest, at least at face value, a need to rethink the galaxy formation physics. For example, by appealing to high star-formation efficiencies of about $10-30\%$ at high redshifts compared to just a few per cent in the local Universe \citep{Moster2013}, a metal-deficient Population~III dominated stellar population that produces more intense UV radiation \citep{Inayoshi2022}, temperature-dependant IMF \citep{Sneppen2022, Steinhardt2022}, or different dark energy models \citep{EDE2022, BK2022, Menci2022}. Alternatively, due to the current lack of spectroscopic confirmations, it might simply point to yet unknown systematic uncertainties like selection effects \citep{Mason2022}, incorrect estimates of the redshift \citep{Finkelstein2022}, or disparate star-formation histories \citep{Endsley2022, Tacchella2022a}. 
\end{enumerate}

One issue that might affect the results presented in this work is the relatively low resolution of the largest-volume MTNG740 simulation. While we have tried to correct for this by scaling up the results to match the highest-resolution TNG50 simulation, this procedure will not fully resolve this problem because the formation of the very first galaxies is still delayed in the low-resolution runs. We hope to overcome this in the near future, by performing higher-resolution, large-volume simulations, but only running them down to $z=10$, to keep the computational cost down. In conclusion, if the results of the early release \jwst observations are spectroscopically confirmed then it might require more sophisticated galaxy formation modelling that takes into account additional physics that so far has not been broadly incorporated into most galaxy formation models. However, we note that any new process invoked to solve this tension must only affect the properties of galaxies at very high redshifts, such that by $z=10$ and below the successful predictions of the fiducial galaxy formation model, which has been tuned to match local Universe observations, are not altered. It is not yet clear which physics modifications can fulfil this non-trivial constraint. We plan to investigate some of theese interesting possibilities in future works.

\section*{Acknowledgements}

We thank the anonymous referee for constructive and insightful comments. We thank Sandro Tacchella, Charlotte Mason, Ryan Endsley and Josh Borrow for helpful discussions and suggestions. The authors gratefully acknowledge the Gauss Centre for Supercomputing (GCS) for providing computing time on the GCS Supercomputer SuperMUC-NG at the Leibniz Supercomputing Centre (LRZ) in Garching, Germany, under project pn34mo. This work used the DiRAC@Durham facility managed by the Institute for Computational Cosmology on behalf of the STFC DiRAC HPC Facility, with equipment funded by BEIS capital funding via STFC capital grants ST/K00042X/1, ST/P002293/1, ST/R002371/1 and ST/S002502/1, Durham University and STFC operations grant ST/R000832/1. CH-A acknowledges support from the Excellence Cluster ORIGINS which is funded by the Deutsche Forschungsgemeinschaft (DFG, German Research Foundation) under Germany’s Excellence Strategy – EXC-2094 – 390783311. VS and LH acknowledge support by the Simons Collaboration on “Learning the Universe”. LH is supported by NSF grant AST-1815978. SB is supported by the UK Research and Innovation (UKRI) Future Leaders Fellowship [grant number MR/V023381/1].

\section*{Data Availability}

The data underlying this article will be shared upon reasonable request to the corresponding authors. All MTNG simulations will be made publicly available in 2024 at \url{www.mtng-project.org}. All \thesan simulation data will be made publicly available in the near future and distributed via \url{www.thesan-project.com}.



\bibliographystyle{mnras}
\bibliography{bibliography} 

\bsp	
\label{lastpage}
\end{document}